\documentclass[modern]{aastex61}

\usepackage{graphicx}
\usepackage{amsmath}
\usepackage{amssymb}
\usepackage{enumitem}

\newcommand\mybar{\kern1pt\rule[-\dp\strutbox]{.8pt}{\baselineskip}\kern1pt}

\setlist[itemize]{noitemsep, topsep=0pt, leftmargin=*}

\shorttitle{Discovery of Close-In Interstellar Objects}
\shortauthors{Loeb}

\begin{document}

\title{Discovering Numerous Interstellar Objects with A Dedicated
  Space Telescope}

\author{Abraham Loeb}

\affiliation{Astronomy Department, Harvard University, 60 Garden St.,
  Cambridge, MA 02138, USA}

\begin{abstract}
  I show that a dedicated space telescope with a meter-size aperture
  can detect numerous interstellar objects, 10-m in diameter, that
  pass within $\sim 20^{\circ}$ from the Sun. Separating the emitted
  thermal radiation from the reflection of sunlight would allow to
  measure the surface temperature, area and albedo of these
  objects. Spectroscopic observations of any evaporated material at
  the expected temperature of $\gtrsim 600~{\rm K}$ would provide
  important clues about the nature and birth sites of interstellar
  objects.
\end{abstract}

\section{Introduction}

The discovery of interstellar objects, such as 1I/`Oumuamua or
2I/Borisov, opened a new path for studying planetary systems around
other stars~\citep{Siraj,Seligman,Jewitt}.

Within our own Solar system, newly-published JWST data from the main
asteroid belt~\citep{Burdanov} reveals a cumulative number of asteroids
larger than a diameter $D$ that scales as a power-law with $D$,
\begin{equation}
  N(>D) \propto D^{q}~,
\label{eq:q}
\end{equation}
with an index of $q=-2.66\pm 0.6$ for $D<100~{\rm m}$.

The first interstellar object, 1I/`Oumuamua with $D\sim 100~{\rm
  m}$~\citep{Meech,Jewitt}, provided a preliminary estimate for the
number density of interstellar objects of its size~\citep{Do},
$n(D>100~{\rm m})\sim 0.1~{\rm au}^{-3}$. Assuming that smaller
interstellar objects follow a distribution similar to that in
equation~(\ref{eq:q}) down to $D\sim 10~{\rm m}$, implies an
interstellar density,
\begin{equation}
  n(D>10~{\rm m})\approx 10^{1.7\pm0.6}~{\rm au}^{-3}~.
  \label{eq:nI}
  \end{equation}

Starting this year, the {\it Legacy Survey of Space and Time} of the
{\it Vera C. Rubin Observatory} will be able to detect interstellar
objects with $D\sim 100~{\rm m}$ within a distance of $\sim 2~{\rm
  au}$~\citep{Ivezic,Seligman1,Siraj1,Seligman2}. The observed flux of
reflected sunlight from objects of a given albedo scales as $\propto
D^2/(d^2r^2)$, where $d$ is their distance from the Sun and $r$ is
their distance from Earth. Therefore, the detectability of objects
with $D\sim 10~{\rm m}$ and $d\sim~0.35~{\rm au}$ (similar to the
semimajor axis of Mercury) is slightly better than that of objects
with $D\sim 100~{\rm m}$ and $r\sim d\sim 2~{\rm au}$.

The outline of this {\it Letter} is as follows. In \S 2, I calculate
the expected arrival and departure rate of 10-m interstellar objects
to a heliocentric distance of $\sim 0.35~{\rm au}$. This distance
corresponds to a transverse angular separation on the sky of $\sim
20^{\circ}$ from the Sun, larger by 2.5 orders of magnitude than the
extent of the field-of-view of the {\it Hubble Space Telescope}
(HST)~\citep{Illingworth}. The main conclusions are summarized in \S
3.

\section{Rate of Passages of Close-In Interstellar Objects}

The arrival and departure rate of interstellar objects with $D\sim
10~{\rm m}$ and a characteristic interstellar velocity $v$ to within a
heliocentric distance $d$ is given by~\citep{Forbes},
\begin{equation}
  {\cal R}(d)= n(D>10~{\rm m})\left\{2\pi d^2 \left[1+{2GM_\odot\over
      dv^2}\right]\right\} v~,
  \label{eq:rate}
  \end{equation}
where the term in square-brackets accounts for gravitational focusing,
which dominates in our regime of interest. Substituting
equation~(\ref{eq:nI}) and $v\sim 30~{\rm km~{\rm s^{-1}}}$ for
interstellar objects originating from the thin disk of stars in the
Milky-Way galaxy~\citep{Forbes}, I get from equation~(\ref{eq:rate}),
\begin{equation}
{\cal R}(d=0.35~{\rm au})\approx 4.4~(\times 10^{\pm 0.6})~{\rm day}^{-1}~.
\end{equation}

Gravitational acceleration increases the characteristic speed of
interstellar objects at a heliocentric distance of $d\sim 0.35~{\rm
  au}$ to a value in the range $\sim (71$--$90)~{\rm km~s^{-1}}$,
implying that their crossing time is of order $\sim (2\times 0.35~{\rm
  au}/80~{\rm km~s^{-1}})\sim 2.2~{\rm weeks}$.

The expected discovery rate is reduced to once per $\sim 1.3~(\times
10^{\pm 0.6})$ weeks within the limb of a restricted $1^\circ$
field-of-view. Given their characteristic speed, interstellar objects
will cross this restricted field-of-view within $\sim 9$ hours -
allowing sufficient time for multiple snapshots. Simultaneous
observations across ${\cal N}$ fields of view, each spanning $\sim
1^\circ$, along a circle of radius $\sim 20^{\circ}$ around the Sun
would increase the detection rate of new interstellar objects by a
factor of ${\cal N}$.

The observed V-band magnitude of reflected sunlight from 1I/`Oumuamua
reached peak values of 22.5 mag at $d\sim 1.43~{\rm au}$ and $R\sim
0.5~{\rm au}$~\citep{Drahus,Mashchenko}. Scaling to an interstellar
object of similar albedo, but with $D\sim 10~{\rm m}$, $d\sim
0.35~{\rm au}$ and $R\sim 1~{\rm au}$, implies a V-band magnitude of
26 mag. Based on the documented performance of
HST~\citep{Illingworth}, objects of this V-band magnitude would be
detectable with a signal-to-noise ratio of 10 in $\sim 3$-hour 
exposures by a space telescope with a meter-size aperture. The
existence of a cometary tail of dust or gas which scatters sunlight
around the object, as observed for 2I/Borisov~\citep{Borisov}, would
improve the detectability prospects of even smaller cores of
interstellar comets.

\section{Discussion}

New interstellar objects of 10-m diameter are expected to enter and
exit a $20^{\circ}$ circle around the Sun once per $\sim 5.5(\times
10^{\pm 0.6})$ hours. HST's Sun-avoidance region is within $50^\circ$
from the
Sun\footnote{\url{https://hst-docs.stsci.edu/hsp/the-hubble-space-telescope-primer-for-cycle-33}}. However,
a future space telescope with a meter-size aperture can be designed to
withstand the excess Solar heat (in the spirit of the examples of the
space-based {\it Solar and Heliospheric
  Observatory}\footnote{\url{https://soho.nascom.nasa.gov/}} and {\it
  Parker Solar
  Probe}\footnote{\url{https://science.nasa.gov/mission/parker-solar-probe/}}
or the ground-based {\it Inouye Solar
  Telescope}\footnote{\url{https://nso.edu/telescopes/inouye-solar-telescope/}}),
and optimize a search strategy that maximizes the discovery rate of
new interstellar objects at even closer-in values of $d$.
Near-twilight observations by the {\it Vera C. Rubin Observatory}
might also detect close-in interstellar objects. The $20^\circ$ circle
extends to $\sim 75R_\odot$, well outside the Solar corona. In
principle, rarer but brighter Sun-divers could also be
identified~\citep{Forbes}.

Interstellar objects can be distinguished from Solar system asteroids
and comets by their orbital speed. Their speed at a perihelion
distance $d$ exceeds the escape speed from the Sun's gravity,
\begin{equation}
v_{\rm esc} =\left({2GM_\odot \over d}\right)^{1/2}= 71.3~{\rm
  km~s^{-1}}\left({d\over 0.35~{\rm au}}\right)^{-1/2}~.
\end{equation}
Their interstellar speed relative to the Sun outside the Solar system,
$v$, can be deduced from their excess speed $\Delta v$ relative to
$v_{\rm esc}$, namely $v=\sqrt{[(\Delta v+v_{\rm esc})^2-v_{\rm
      esc}^2]}$.

At a perihelion distance $d$, the effective Sun-facing surface
temperature of the interstellar objects,
\begin{equation}
T\approx 600~{\rm K} \left({d\over 0.35{\rm au}}\right)^{-1/2}~,
\end{equation}
corresponds to a blackbody peak-brightness at a wavelength of
\begin{equation}
\lambda_{\rm peak}\approx 4.8\mu{\rm m}~\left({d\over 0.35{\rm
    au}}\right)^{1/2}~.
\end{equation}

Separating the two blackbody components of reflected sunlight
($T_\odot\sim 5,800~{\rm K}$) and emitted thermal radiation ($T\sim
600~{\rm K}$) would allow to determine the diameter of the
interstellar objects, $D\approx (2L/\pi\sigma T^4)^{1/2}$, based on
their surface temperature $T$ and total emitted luminosity $L$ from
the hotter Sun-facing hemisphere, where $\sigma$ is the
Stefan-Boltzmann constant. Knowing their surface area and reflected
sunlight flux would determine their albedo for sunlight.

At a surface temperature of $T\sim 600~{\rm K}$, some interstellar
objects are likely to evaporate and potentially
disintegrate. Spectroscopy of their cometary tail~\citep{Forbes} and
non-gravitational acceleration~\citep{Micheli} would reveal their
chemical composition and provide important clues about their nature
and potential birth sites~\citep{Morgan}. These diagnostics would
clarify whether the anomalous non-gravitational acceleration of
interstellar objects like 1I/`Oumuamua resembles that of dark comets
in the Solar system~\citep{Seligman3}.

\bigskip
\bigskip
\bigskip
\bigskip
\section*{Acknowledgements}

This work was supported in part by the {\it Galileo Project} at
Harvard University. I thank Frank Laukien for insightful discussions
that inspired this work and Fabio Pacucci and Sriram Elango for
helpful comments.
 
\bigskip
\bigskip
\bigskip

\bibliographystyle{aasjournal}
\bibliography{t}

\begin{thebibliography}{}
\expandafter\ifx\csname natexlab\endcsname\relax\def\natexlab#1{#1}\fi
\providecommand{\url}[1]{\href{#1}{#1}}
\providecommand{\dodoi}[1]{doi:~\href{http://doi.org/#1}{\nolinkurl{#1}}}
\providecommand{\doeprint}[1]{\href{http://ascl.net/#1}{\nolinkurl{http://ascl.net/#1}}}
\providecommand{\doarXiv}[1]{\href{https://arxiv.org/abs/#1}{\nolinkurl{https://arxiv.org/abs/#1}}}

\bibitem[{{Bodewits} {et~al.}(2020){Bodewits}, {Noonan}, {Feldman},
  {Bannister}, {Farnocchia}, {Harris}, {Li}, {Mandt}, {Parker}, \&
  {Xing}}]{Borisov}
{Bodewits}, D., {Noonan}, J.~W., {Feldman}, P.~D., {et~al.} 2020, Nature
  Astronomy, 4, 867, \dodoi{10.1038/s41550-020-1095-2}

\bibitem[{{Burdanov} {et~al.}(2025){Burdanov}, {de Wit}, {Bro{\v{z}}},
  {M{\"u}ller}, {Hoffmann}, {Ferrais}, {Micheli}, {Jehin}, {Parrott}, {Hasler},
  {Binzel}, {Ducrot}, {Kreidberg}, {Gillon}, {Greene}, {Grundy}, {Kareta},
  {Lagage}, {Moskovitz}, {Thirouin}, {Thomas}, \& {Zieba}}]{Burdanov}
{Burdanov}, A.~Y., {de Wit}, J., {Bro{\v{z}}}, M., {et~al.} 2025, \nat, 638,
  74, \dodoi{10.1038/s41586-024-08480-z}

\bibitem[{{Do} {et~al.}(2018){Do}, {Tucker}, \& {Tonry}}]{Do}
{Do}, A., {Tucker}, M.~A., \& {Tonry}, J. 2018, \apjl, 855, L10,
  \dodoi{10.3847/2041-8213/aaae67}

\bibitem[{{Drahus} {et~al.}(2018){Drahus}, {Guzik}, {Waniak}, {Handzlik},
  {Kurowski}, \& {Xu}}]{Drahus}
{Drahus}, M., {Guzik}, P., {Waniak}, W., {et~al.} 2018, Nature Astronomy, 2,
  407, \dodoi{10.1038/s41550-018-0440-1}

\bibitem[{{Forbes} \& {Loeb}(2019)}]{Forbes}
{Forbes}, J.~C., \& {Loeb}, A. 2019, \apjl, 875, L23,
  \dodoi{10.3847/2041-8213/ab158f}

\bibitem[{{Hoover} {et~al.}(2022){Hoover}, {Seligman}, \& {Payne}}]{Seligman1}
{Hoover}, D.~J., {Seligman}, D.~Z., \& {Payne}, M.~J. 2022, PSJ, 3, 71,
  \dodoi{10.3847/PSJ/ac58fe}

\bibitem[{{Illingworth} {et~al.}(2013){Illingworth}, {Magee}, {Oesch},
  {Bouwens}, {Labb{\'e}}, {Stiavelli}, {van Dokkum}, {Franx}, {Trenti},
  {Carollo}, \& {Gonzalez}}]{Illingworth}
{Illingworth}, G.~D., {Magee}, D., {Oesch}, P.~A., {et~al.} 2013, \apjs, 209,
  6, \dodoi{10.1088/0067-0049/209/1/6}

\bibitem[{{Ivezi{\'c}} \& {Ivezi{\'c}}(2021)}]{Ivezic}
{Ivezi{\'c}}, V., \& {Ivezi{\'c}}, {\v{Z}}. 2021, \icarus, 357, 114262,
  \dodoi{10.1016/j.icarus.2020.114262}

\bibitem[{{Jewitt}(2024)}]{Jewitt}
{Jewitt}, D. 2024, arXiv e-prints, arXiv:2407.06475,
  \dodoi{10.48550/arXiv.2407.06475}

\bibitem[{{Jewitt} \& {Seligman}(2023)}]{Seligman}
{Jewitt}, D., \& {Seligman}, D.~Z. 2023, \araa, 61, 197,
  \dodoi{10.1146/annurev-astro-071221-054221}

\bibitem[{{Loeb} \& {MacLeod}(2024)}]{Morgan}
{Loeb}, A., \& {MacLeod}, M. 2024, A\&A, 686, A123,
  \dodoi{10.1051/00004-6361/202449250}

\bibitem[{{Mar{\v{c}}eta} \& {Seligman}(2023)}]{Seligman2}
{Mar{\v{c}}eta}, D., \& {Seligman}, D.~Z. 2023, PSJ, 4, 230,
  \dodoi{10.3847/PSJ/ad08c1}

\bibitem[{{Mashchenko}(2019)}]{Mashchenko}
{Mashchenko}, S. 2019, \mnras, 489, 3003, \dodoi{10.1093/mnras/stz2380}

\bibitem[{{Meech} {et~al.}(2017){Meech}, {Weryk}, {Micheli}, {Kleyna},
  {Hainaut}, {Jedicke}, {Wainscoat}, {Chambers}, {Keane}, {Petric}, {Denneau},
  {Magnier}, {Berger}, {Huber}, {Flewelling}, {Waters}, {Schunova-Lilly}, \&
  {Chastel}}]{Meech}
{Meech}, K.~J., {Weryk}, R., {Micheli}, M., {et~al.} 2017, \nat, 552, 378,
  \dodoi{10.1038/nature25020}

\bibitem[{{Micheli} {et~al.}(2018){Micheli}, {Farnocchia}, {Meech}, {Buie},
  {Hainaut}, {Prialnik}, {Sch{\"o}rghofer}, {Weaver}, {Chodas}, {Kleyna},
  {Weryk}, {Wainscoat}, {Ebeling}, {Keane}, {Chambers}, {Koschny}, \&
  {Petropoulos}}]{Micheli}
{Micheli}, M., {Farnocchia}, D., {Meech}, K.~J., {et~al.} 2018, \nat, 559, 223,
  \dodoi{10.1038/s41586-018-0254-4}

\bibitem[{{Seligman} {et~al.}(2024){Seligman}, {Farnocchia}, {Micheli},
  {Hainaut}, {Hsieh}, {Feinstein}, {Chesley}, {Taylor}, {Masiero}, \&
  {Meech}}]{Seligman3}
{Seligman}, D.~Z., {Farnocchia}, D., {Micheli}, M., {et~al.} 2024, Proceedings
  of the National Academy of Science, 121, e2406424121,
  \dodoi{10.1073/pnas.2406424121}

\bibitem[{{Siraj} \& {Loeb}(2022)}]{Siraj}
{Siraj}, A., \& {Loeb}, A. 2022, Astrobiology, 22, 1459,
  \dodoi{10.1089/ast.2021.0189}

\bibitem[{{Siraj} {et~al.}(2023){Siraj}, {Loeb}, {Moro-Mart{\'\i}n}, {Elowitz},
  {White}, {Watters}, {Melnick}, {Cloete}, {Grindlay}, \& {Laukien}}]{Siraj1}
{Siraj}, A., {Loeb}, A., {Moro-Mart{\'\i}n}, A., {et~al.} 2023, Journal of
  Astronomical Instrumentation, 12, 2340001, \dodoi{10.1142/S2251171723400019}

\end{thebibliography}
\label{lastpage}
\end{document}